\documentclass[prd,aps,twocolumn,showpacs,preprintnumbers,amsmath,amssymb]{revtex4}
\usepackage[dvips]{graphicx} 
\usepackage{epsf}
\usepackage{amsmath} 
\usepackage{amssymb} 
 
\usepackage{graphicx}
\usepackage{dcolumn}
\usepackage{bm}
\def\be{\begin{equation}} 
\def\ee{\end{equation}} 
\def\bea{\begin{eqnarray}} 
\def\eea{\end{eqnarray}} 

\newcommand{\comment}[1]{}

\newcommand{\tphi}{\tilde{\phi}}

\begin{document} 
 
 
\date{\today} 
 
\title{Cosmic Ray Positrons from Cosmic Strings} 
 
\author{Robert Brandenberger$^{1,2,3}$, Yi-Fu Cai$^{2}$, Wei Xue$^{1}$,  and Xinmin Zhang$^{2,3}$} 
 
\affiliation{1) Department of Physics, McGill University, 
Montr\'eal, QC, H3A 2T8, Canada} 
 
\affiliation{2) Institute of High Energy Physics, Chinese Academy of Sciences, P.O. Box 918-4, 
Beijing 100049, P.R. China}

\affiliation{3) Theoretical Physics Center for Science Facilities (TPCSF), Chinese Academy of
Sciences, P.R. China}

\pacs{98.80.Cq} 
 
\begin{abstract} 

We study the spectrum of cosmic ray positrons produced by a scaling distribution
of non-superconducting cosmic strings. In this scenario, the positrons are produced
from the jets which form from the cosmic string cusp annihilation process.
The spectral shape is a robust feature of our scenario, and is in good agreement
with the results from the recent PAMELA and ATIC experiments. In particular, the
model predicts a sharp upper cutoff in the spectrum, and a flux which rises
as the upper cutoff is approached. The energy at which the flux peaks is determined
by the initial jet energy. The amplitude of the flux can be adjusted
by changing the cosmic string tension and also depends on the cusp annihilation
efficiency.
  
\end{abstract} 
 
\maketitle

\newcommand{\eq}[2]{\begin{equation}\label{#1}{#2}\end{equation}} 
 
\section{Introduction} 

Recent results from the PAMELA \cite{Pamela} and ATIC \cite{Atic} experiments
have indicated an excess power of the cosmic ray positron flux  compared to
what is predicted from astrophysical backgrounds alone. The power of the
flux is observed to rise towards an upper cutoff which is in the range of $600$GeV
and falls off quite sharply above this cutoff. The data is in conflict
with what is expected from astrophysical backgrounds. A peak in the
spectrum of cosmic ray positrons in the energy range of $80 - 300$GeV
had been suggested as a signature of dark matter annihilation
in the galaxy a long time ago (the positrons are either produced from 
jets formed by the non-leptonic decay products \cite{Silk,Stecker,Ellis} 
or by direct decay into electrons and positrons \cite{TW,KaT,Tylka} - 
for reviews see e.g. \cite{Jungman,Bertone}). However, the specific
form of the spectrum obtained from observations is hard to
reconcile with the predictions from simple dark matter annihilation models
\cite{challenge}, although modified models have been proposed which
are in better agreement with the data \cite{modified}. It is also possible
that the positrons are due to nearby pulsars \cite{pulsarexpl}. However,
in the case of pulsars the decline of the flux at energies larger
than the peak energy is not abrupt (see, however, \cite{Ioka} for a
different point of view on this issue).

In this Letter we would like to propose an alternative explanation for
the positron excess which has nothing to do with dark matter. We
investigate the possibility that the observed positron flux is due to
jets from cusp annihilation of cosmic string loops. We are considering
non-superconducting cosmic strings. It has been known for a long time
that particle emission \cite{RHBcusp} from cusps of cosmic strings leads 
to a spectrum which rises as a
function of energy up to a cutoff set by the parameters in the cosmic string
model and falls off quite sharply above this cutoff \cite{Jane}. 

Since cosmic strings arise in a large class of particle physics beyond
the Standard Model, our mechanism provides a way to test physics
beyond the Standard Model independent of the existence of low-energy
supersymmetry.

In the following section we review the basics of cosmic string dynamics
and particle emission from cosmic strings which are important to
understand our scenario. Section 3 contains the computation of
the cosmic ray positron flux. In Section 4 we give a brief
discussion of the change in the positron spectrum during propagation
through the galaxy. We conclude with a discussion of our
results.
 
\section{Particle Emission from a Cosmic String Loop}

In this section we review the basics of cosmic string dynamics and
discuss the cusp annihilation mechanism by which cosmic string
loops can emit high energy particles.

Cosmic strings are one-dimensional topological defects which
form during symmetry breaking phase transitions 
in a wide class of gauge theory models 
(see \cite{ShellVil,HK,RHBrev}  for reviews). 
If the gauge symmetry group at high temperatures is $G$, and
the unbroken subgroup below the transition temperature is $H$,
then the criterion for the existence of cosmic strings in the theory
is
\be
\Pi_1 (\cal{M}) \, \neq \, \cal{I} \, ,
\ee
where $\cal{M}$ is the vacuum manifold of the theory below
the transition temperature and $\cal{I}$ is the trivial group. 
If the group $G$ is simply connected,
then ${\cal M} = G/H$. This criterion is satisfied in a large class
of particle physics theories beyond the Standard Model. 

In particle physics theories admitting the existence of cosmic
strings, such strings inevitably arise during the symmetry
breaking phase transition \cite{Kibble}. By causality,
the point in $\cal{M}$ which the order parameter describing the
phase transition takes at temperatures lower than the transition
temperature must be uncorrelated on length scales larger
than the Hubble radius $H^{-1}$, where $H$ is the
cosmological expansion rate. Hence, there is a probability
of order 1 that one string will traverse any particular Hubble volume
after the phase transition. Typically (in particular if matter
above the transition temperature is in thermal equilibrium),
the initial separation of the cosmic strings will be microscopic.
The above causality argument applies at all times subsequent
to the transition time. Thus, in a theory which admits cosmic
strings, then at all times after the phase transition (in particular
at recent cosmological times) a network of cosmic strings
with separation no greater than the Hubble radius will be present.

Cosmic strings arising in gauge field theories must be closed, i.e.
either string loops of ``infinite" strings (defined as truly
infinite strings or string loops with curvature radius larger than
the Hubble radius. Thus,
the system of cosmic strings at any time after the phase
transition will consist of a network of ``infinite" strings 
and a distribution of string loops. Both analytical
arguments detailed in \cite{ShellVil,HK,RHBrev}  and
detailed numerical simulations \cite{AT,BB,AS}
have shown that the network of infinite strings approaches a
``scaling" solution characterized by a string correlation length
which is a fixed fraction of the Hubble radius at all late times.
Roughly speaking, we can view the long string network as
a random walk with step length of the order of the 
Hubble radius.

The scaling solution implies that the total length in the long
strings is decreasing. This decrease is realized by the
inter-commutation of long strings. Such inter-commutations
produce string loops. Thus, at any time $t$, there will
be distribution of loops. Neither numerical simulations nor
analytical studies at this point
agree on the exact nature of the loop distribution (see e.g.
\cite{Pol} for recent progress). We will be using a simple
one-scale model for the distribution of loops which is based
on the assumption that loops at time $t$ form at a fixed
fraction of the Hubble radius. 

Once formed, string loops decay predominantly by
gravitational radiation. The rate of gravitational
radiation is governed by the string tension $\mu$, namely
\be \label{decay}
{\dot R} \, = \, \gamma G \mu \, ,
\ee
where $G$ is Newton's gravitational constant and $\gamma$
is a numerical constant whose value is of the order $10^2$
\cite{CSgwrad}.

Taking into account the redshift of the number density of
string as well as the formation scenario and decay rate of
cosmic strings discussed above, we obtain the following
distribution of string loops at time $t$ (see e.g. \cite{BT}):
\bea \label{loopdistrib}
n(R, t) \, &=& \, \kappa R^{-2} t^{-2} \,\, , \,\, R \, > \, t_{eq} \, \\
n(R, t) \, &=& \, \kappa R^{-5/2} t^{-2} t_{eq}^{1/2} \,\, , \,\,  
\gamma G \mu t \, < \, R \, < \, t_{eq} \, \nonumber \\ 
n(R, t) \, &=& \, \kappa (\gamma G \mu)^{-5/2} t^{-2} t_{eq}^{1/2} \,\, , \,\,  
 R \, < \, \gamma G \mu t \, , \nonumber
\eea
where $\kappa$ is another numerical constant which depends on the
details of the cosmic string scaling solution,
and $t_{eq}$ is the time of equal matter and radiation. In the above, we
are assuming $\gamma G \mu \, < \, t_{eq}$. The first line in
(\ref{loopdistrib}) represents loops which were produced after the
time of equal matter and radiation, the second line loops generated
before $t_{eq}$ which will survive gravitational radiation for more
than a Hubble expansion time. The last line represents loops
which are in the final stages of decay by gravitational radiation.

From the distribution (\ref{loopdistrib}) it can be seen that 
the energy density in string loops is dominated by loops of radius
about $R \, \sim \, \gamma G \mu t$. These loops will, as we show
below, also dominate the positron emission from strings.

Cosmic strings as two-dimensional world sheets $x^{\mu}(\sigma, \tau)$
(where $\tau$ is a world sheet time coordinate and $\sigma$ labels
the spatial world sheet coordinate) are solutions of
the Nambu-Goto equations, the same equations
which describe fundamental strings. Since cosmic strings
have relativistic tension, they will oscillate. It can be shown \cite{KT}
that ``cusps" generically occur on string loops (at least once per
oscillation time). A cusp is a point on the string where 
\be
x_{i}^{\prime} \, = \, 0 \, ,
\ee
where a prime indicates the derivative with respect to $\sigma$.

Geometrically, a cusp corresponds to a ``spike" on the string (see
Figure 1). Since a cosmic string has a finite width $w$ whose magnitude
is of the order of $\mu^{-1/2}$, at a cusp the two segments of
the string at either side of the cusp overlap. By expanding the
solutions of the Nambu-Goto equations about a cusp it can be
shown \cite{Spergel} that the length of the overlap region is
\be
l_c \, \sim \, w^{1/3} R^{2/3} \, ,
\ee
where $R$ is the radius of the string loop. 

\begin{figure}[htbp]
\includegraphics[scale=0.6]{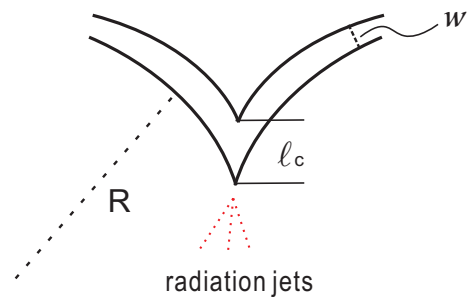}
\caption{Sketch of a cusp on a cosmic string loop. The mean
curvature radius of the string loop is $R$, the string width is $w$,
and the length of the overlap region is $l_c$.}
\end{figure}

There is nothing that prevents this overlap region from annihilating 
into particle excitations of the Higgs and gauge fields which the
string is made up of. Assuming that the entire overlap region
decays, one gets a power of particle radiation given by \cite{RHBcusp}
(see also \cite{RHBcusp2})
\be  \label{cuspemission}
P_c \, \sim \, \mu l_c R^{-1} \, \sim \, \mu w^{1/3} R^{-1/3} \, .
\ee
This particle decay rate is - for string loops of macro-physical radius
$R$ - very small compared to the power radiated into gravitational
radiation which is
\be
P_g \, = \gamma (G \mu) \mu 
\ee
(see (\ref{decay})). Nevertheless, as was discussed in \cite{Jane}
(see also \cite{Bhatta}),  cusp annihilation can contribute a
significant fraction to the cosmic ray flux, assuming that a
substantial fraction of the string overlap region decays \footnote{Note
that the back-reaction of one string segment on the other may
in fact prevent the cusp from ever reaching its full extent $l_c$
\cite{Olum}. This will lead to a substantial reduction of the cusp
emission. In the following we will include this effect via an
``efficiency factor" $\varepsilon$, a numerical coefficient smaller
than 1 which will multiply the expression on the right-hand side
of (\ref{cuspemission}).}

The expected ultra-high energy neutrino flux from cosmic strings
with a tension given by the scale of Grand Unification was studied
in \cite{Jane,Bhatta}, and the corresponding $\gamma$-ray
signatures were analyzed in \cite{Jane2}. These works were
performed under the assumption that the distribution of cosmic
string loops scales as described by (\ref{loopdistrib}). The analysis
was extended in \cite{Jane3} to the case of a non-scaling loop
distribution (numerical evidence for such a non-scaling loop
distribution came from the analysis of \cite{VHS}). However, most
studies of cosmic string dynamics favor a scaling distribution of
string loops (see e.g. \cite{Pol}), and thus in this paper we will
assume such a scaling distribution.

\section{Positron Flux from a Scaling Network of Cosmic Strings}

As discussed above, the power of energy loss from a cosmic string
loop of radius $R$ is
\be
P_c \, = \, \varepsilon \mu^{5/6} R^{-1/3} \, ,
\ee
where $\varepsilon$ is the efficiency factor discussed above and
we have made use of the fact that the string width $w$ is
proportional to the inverse square root of the string tension $\mu$.

The primary decay products from cusp annihilation are quanta
of the scalar and gauge fields which make up the cosmic strings.
These quanta, in turn, will decay into relativistically moving
standard model particles which will form jets. Following the
discussion in \cite{Jane}, we take the primary energy of a single
jet to be $m_f$. In this case, the number of jets $\dot{N}$
formed per unit time is
\be \label{Pdef}
{\dot N} \, = \, \varepsilon \mu^{5/6} m_f^{-1} R^{-1/3} \, \equiv \, P R^{-1/3}\, ,
\ee
where the last step defines the quantity $P$. 

A single jet leads to the following spectrum of energies 
(number per energy interval) of neutrinos resulting from the jet \cite{jet}:
\be \label{jetspec}
\frac{dN}{dE} \, = \, \frac{15}{16} m_f^{-1}
\bigl( \frac{11}{3} - 6 x^{1/2} - 4 x^{-1/2} + \frac{2}{3} x^{-3/2} \bigr) \, , 
\ee
where
\be
x \, \equiv \, \frac{E}{m_f} \, < \, 1 \, .
\ee
We will take the same formula to give the energy spectrum of all stable
leptons resulting from the decay, in particular the spectrum of
positrons.

Particle physics models admitting non-superconducting cosmic string solutions
will have string cusps decaying at all times. Neutrinos produced from
cusp decay will travel cosmological distances (see e.g. \cite{Jane}),
whereas positrons are absorbed and lose their energy on super-galactic
scales. To obtain the neutrino flux at energy $E$, we have to integrate over all 
times $t$ the flux of particles emitted at time $t$ with energy $E (z(t) + 1)$. For
electrons and positrons, we only have to integrate over times which are
smaller than the current time $t_0$ by less than the ``containment time"
of electrons and positrons, respectively, in the galaxy. According to
\cite{TW}, the containment time of positrons is of the order
$10^{7}$ yrs,, i.e. much longer than the 
time it would take light to travel through the galaxy. The containment time
corresponds to a redshift of $z_c \sim 10^{-3}$.  

When computing the expected positron flux from cosmic string cusp 
annihilations we must therefore impose several cutoffs. First of all,
only positrons emitted at redshifts smaller than $z_c$ will contribute.
Secondly, only string loops located inside the galaxy may be
considered.

The general expression for the differential energy flux $F(E)$ of 
cosmic ray positrons from cosmic string cusp annihilations is
\be \label{flux1}
F(E) \, = \, \int dt (z(t) + 1)^{-3} f((z(t) + 1)E, t) \, ,
\ee
where $f$ is the differential flux per unit time of positrons emitted
at time $t$, the ``injection spectrum". The redshift enters in two places. 
Firstly, the injection
number density is redshifted, and, secondly, the energy of
a given positron redshifts.

The injection spectrum is obtained by integrating over all cosmic
string loops present:
\bea \label{injection}
f((z(t) + 1)E, t) \, &=& \, (z(t) + 1) \frac{dN}{dE}|_{x = (z+1)E/m_f} 
\nonumber \\
& & \, \int dR n(R, t) P R^{-1/3} \, ,
\eea
where the string number density $n(R, t)$ and the constant $P$
have been defined previously. The first factor on the right-hand
side of this equation is the Jacobean factor obtained by
transforming between final energy $E$ and injection energy.

As follows from recalling the loop distribution (\ref{loopdistrib}),
the integral over $R$ is dominated by loops of radius
$R \sim \gamma G \mu t$. If we consider the string scale to
be at least a couple of orders of magnitude smaller than the
scale of Grand Unification, string loops of radius 
$R \sim \gamma G \mu t_0$ will still be present today,
and their number density will be such that many string
loops of such radius will be located within our galaxy.

To estimate the amplitude and shape of the spectrum, we
first insert the injection flux (\ref{injection}) into {\ref{flux1})
and perform the integral over loop radii. The integral
is dominated by the value $R = \gamma G \mu t$. A good
estimate of the result is obtained by
integrating over the loops with radii in the range 
$\gamma G \mu t \, < \, R \, < \, t_{eq}$. The result is
\bea
F(E) \, &\sim& \, P  \nu (\gamma G \mu)^{-11/6} t_{eq}^{1/2} \\
& & \, \int dt (z(t) + 1)^{-2} t^{-2} t^{-11/6} \frac{dN}{dE}|_{x = (z+1)E/m_f}  \, .
\nonumber
\eea
Next is the integral over time which can be simplified by using
the integration variable
\be
{\tilde z} \, \equiv \, z(t) + 1 \, .
\ee
To obtain an estimate of the flux, we use for $dN/dE$ the final
term on the right-hand side of (\ref{jetspec}). After a couple of
lines of algebra (and in particular plugging in the definition of 
$P$ from(\ref{Pdef})) we obtain
\bea \label{flux2}
E^3 F(E) \, &\sim& \, \varepsilon \nu (\gamma G \mu)^{-11/6} z_{eq}^{-3/4} \\ 
& & \, t_0^{-7/3} \mu^{5/6} m_f \bigl( \frac{E}{m_f}\bigr)^{3/2} z_c \, , \nonumber
\eea
where the final factor comes from the range of integration over $t$.

Inserting numbers into (\ref{flux2}) and expressing the result in
terms of the units which experimentalists use we get
\bea
E^3 F(E) \, &\sim& \, \varepsilon \nu \gamma^{-5/6} (\gamma G \mu)^{-1} \\
& & \, m_f|_{\rm{GeV}} \bigl( \frac{E}{m_f}\bigr)^{3/2} z_c 10^{-11} m^{-2} \rm{sec}^{-1} \rm{GeV}^2 \, .
\nonumber
\eea

The specific signature of our predicted cosmic ray positron flux is the power law
increase of $E^3 F(E) \sim E^{3/2}$ and the sharp cutoff at an energy scale set
by the initial jet energy $m_f$. In contrast, the background flux of positrons 
(multiplied by $E^3$) from astrophysical sources is predicted to be slightly decreasing
in the energy range between $10$GeV and $1000$GeV.

Let us first give a rough analytical treatment of the predicted positron to electron
flux ratio. Both fluxes are a superposition of background and cosmic string-induced
fluxes, and we will use the subscripts $bg$ and $cs$, respectively, to denote these
two contributions. The flux ratio $\Phi$ is
\be
\Phi \, \equiv \, \frac{E^3 F(E)^{+}}{E^3 F(E)^{-} + E^3 F(E)^{+}} \, .
\ee
Assuming that the electron flux is dominated by the background, we obtain
\be
\Phi \, = \, R_{bg} + \frac{E^3 F(E)^{+}_{cs}}{E^3 F(E)^{-}_{bg}} \,
\ee
where $R_{bg}$ is the background flux ratio. Since $E^3 F(E)^{-}_{bg}$ is
roughly constant, we see that the flux ratio in the cosmic string model
is predicted to be equal to the background value at low energies and
gradually shift to scaling as
\be
\Phi \, \sim \, E^{3/2} 
\ee
at higher energies. At energies close to the cutoff value $m_f$, the
spectrum again flattens out because terms in (\ref{jetspec})
scaling differently than the $x^{-3/2}$ term which we focused
on will become important. 

From the PAMELA data \cite{Pamela} we can read off a slope
which is rising to about $0.5$ at energies between $50$ and $100$GeV.
From the ATIC data, a slope of close to $1$ is inferred at energy
scales between $300$ and $600$GeV.

To obtain a better idea of the fit of our model, we have evaluated
the predicted positron flux numerically, keeping all of the terms
in (\ref{jetspec}). Our results are plotted in Figures 2 - 4. The
numerical results also include the processing of the spectrum
during propagation as discussed in the following section.

\section{Propagation of the Positrons}

Positrons will lose energy not only because of red-shifting, but also
because of interactions during their propagation from the source to
us through the galaxy. We consider a standard diffusion model for the 
propagation of positrons in the galaxy.

To begin with, let us recall the physical processes which 
affect the propagation of charged particles in the galaxy. Firstly,
when a charged particle travels through the galaxy, its movement
can be affected by the galactic magnetic field. Although the
magnetic gyro-radius of a particle is usually very small, this
particle can still possible to jump to near-bye field lines due to
the tangled magnetic field and so could change its orbit. We
usually model this process with a diffusion equation. Secondly,
during the propagation of a positron, the particle loses
energy because of inverse Compton and synchrotron
processes. These two factors are the most important ones. 
For a detailed study, we refer to Ref.
\cite{Strong:1998pw}. In the following we focus on the 
above-mentioned two processes, especially the energy loss

Neglected other effects which are present in addition to the
two mentioned in the previous paragraph,  and assuming a spherically
symmetric diffusion process, we obtain the following
propagation equation for the flux $F$ of charged cosmic ray particles:
\begin{eqnarray}\label{DLeq}
\frac{\partial}{\partial t}F \, = \, D(\epsilon)\nabla^2F +
\frac{\partial}{\partial\epsilon}\bigg(L(\epsilon)F\bigg) +
Q(\epsilon, \vec{x})~,
\end{eqnarray}
where $\epsilon$ is defined as a dimensionless energy variable
$\epsilon\equiv \frac{E}{a~GeV}$ with $a$ the scale factor of the
universe, $D$ is the diffusion coefficient, $L$ is the energy loss
rate and $Q$ is the source term.

In models in which dark matter annihilation is the source of
the positron excess, the production
of positrons is dominated by the annihilation of dark matter particles
today. Therefore, usually only steady state solutions of Eq. (\ref{DLeq}) 
(in which he left hand side of the equation vanishes) are
considered, as analyzed  for example in Ref.\cite{Baltz:1998xv}.

However, in our model the source of positrons does not
scale in time as the background density, and hence the
resulting flux will not be steady state-like.
As discussed in the previous section, we need to
integrate the flux equation from the earliest moment
from which positrons will still reach us today. This time is
the containment time of positrons in the galaxy
which corresponds to a redshift $z_c\sim10^{-3}$.

In the following, we will neglect the diffusion term in (\ref{DLeq}).
In models with a string tension significantly smaller than that
given by the scale of Grant Unification, the separation of
strings is much smaller than the radius of the galaxy. Many of
these strings are a distance away from us which is smaller than
the diffusion radius. Hence, we argue that we can treat the
flux as quasi-homogeneous and hence neglect the
diffusion term.

Thus, we consider the simplified propagation equation
\begin{eqnarray}
\frac{\partial}{\partial t}F \, \simeq \,
\frac{\partial}{\partial\epsilon}\bigg(L(\epsilon)F\bigg)+(z(t)+1)^{-3}f((z(t)+1)E,t)~.
\end{eqnarray}
To solve this equation, we apply a perturbative approach and separate the flux 
$F$ into infinitely many components,
\begin{eqnarray}
F \, = \, F_0+F_1+...~.
\end{eqnarray}
Each component satisfies its own propagation equation as follows,
\begin{eqnarray}
\frac{\partial}{\partial t}F_0 \, &=& \, (z(t)+1)^{-3}f((z(t)+1)E,t)~,\nonumber\\
\frac{\partial}{\partial t}F_1 \, &=& \, \frac{\partial}{\partial\epsilon}\bigg(L(\epsilon)F_0\bigg)~,~...
~\nonumber
\end{eqnarray}
and, more generally,
\begin{eqnarray}
\frac{\partial}{\partial t}F_i \, &= \, &\frac{\partial}{\partial\epsilon}\bigg(L(\epsilon)F_{i-1}\bigg)~,~...~.
\end{eqnarray}
After solving these equations one by one, we then sum up all the
component to obtain the result
\begin{eqnarray}
F(\epsilon) \, = \, \int_{t_0}^{t_i}dt a^3f + \int_{t_0}^{t_i}dt
\frac{\partial}{\partial\epsilon}L(\epsilon)\int_{t_0}^{t}dt'
a^3f+~...~.
\end{eqnarray}
By changing the time integrals to all the entire integral range, we
can obtain the factor $1/n!$ for the $n$-th component. Eventually,
we obtain the following formal solution for the flux:
\begin{eqnarray}\label{formalF}
F(\epsilon) \, = \, F_0(\epsilon,t_0)\times\exp\bigg\{
\frac{\int_{t_0}^{t_i}dt\frac{\partial}{\partial\epsilon}L(\epsilon)F_0(\epsilon,t)}{F_0(\epsilon,t_0)}
\bigg\}~,
\end{eqnarray}
where $F_0(\epsilon,t_0)$ is exactly what we have studied in Sec.
III without considering the energy losing effect.

In a realistic model, we consider energy loss through synchrotron
emission and inverse Compton scattering. As introduced in Ref.
\cite{Longair:1994wu}, the process can be parameterized as
\begin{eqnarray}
L(\epsilon) \, = \, \frac{\epsilon^2}{\tau_E}~,
\end{eqnarray}
with the energy-loss time $\tau_E\simeq10^{16}s$. After combining Eq.
(16) and the energy-loss parametrization, we can derive the factor
in the exponential term of Eq. (\ref{formalF}). It is
\begin{eqnarray}\label{factor}
&\sim& \, \int_{z_0}^{z_0+z_c} dz \frac{3t_0}{4\tau_E}
(z^{-\frac{3}{2}}-z^{-\frac{3}{4}})
E^{-\frac{1}{2}}\bigg/z_cE^{-3/2} \nonumber\\
&\simeq& \, -\frac{9t_0}{32\tau_E}\frac{z_c E}{1GeV}~,
\end{eqnarray}
in the low energy regime. Correspondingly, we obtain the
following approximate form of the flux
\begin{eqnarray}
F(E) \, \simeq \, F_0(E)\exp\{-\frac{9t_0}{32\tau_E}\frac{z_cE}{1GeV}\}~.
\end{eqnarray}
The exponential factor describes the energy
losing of the positrons when they are passing through the galaxy.

From the above result, we learn that because of the smallness of
the containment time of positrons in the galaxy, the energy
loss due to interactions is insignificant for low energy positrons.
For higher energy positrons the energy loss becomes more important.
This leads to a slight smoothing out of the delta function-like upper
cutoff in the predicted flux. In addition, depending on the parameters,
the energy corresponding to the maximum of the flux may be smaller
than $m_f$. 

To obtain a better idea of the fit of our model, we have evaluated
the predicted positron flux numerically, keeping all of the terms
in (10) and taking into account of the energy loss effects on the
propagation of positrons discussed in this section. 
Our results are plotted in Figures \ref{fig:flux},
\ref{fig:pamela} and \ref{fig:ratio}. In the figures, we take
three groups of parameters for the model as shown in the captions
of these figures. We have chosen the parameters $\nu=13$
(determined by numerical simulations of cosmic string evolution
\cite{AS}), the containment time $z_c=10^{-3}$, and the
energy-loss time $\tau_E=10^{16}s$.

\begin{figure}[htbp]
\includegraphics[scale=0.7]{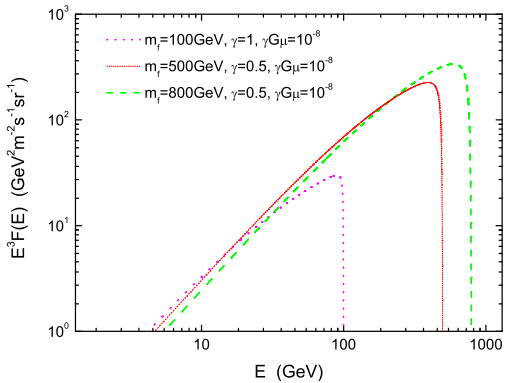}
\caption{The predicted positron flux from cosmic string cusp decay for
various values of the string tension $G \mu$ and the initial jet mass of 
$m_f $. The values of the other parameters were chosen to be 
$\nu = 13$ and efficiency factor $\epsilon = 1$.}
\label{fig:flux}
\end{figure}

\begin{figure}[htbp]
\includegraphics[scale=0.7]{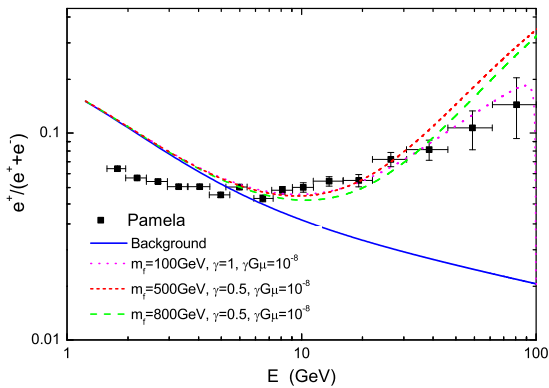}
\caption{Comparison of the cosmic string model predictions
for the flux ratio $\Phi$ with the data from the PAMELA
experiment, for the same choices of parameters as in the previous
figure.}
\label{fig:pamela}
\end{figure}

Whereas the PAMELA experiment only shows a rise of the
flux ratio as a function of energy, the ATIC experiment which probes
the spectrum of positrons to higher energies shows a sharp
upper cutoff at an energy of about $600$ GeV. Thus, matching
the ATIC data leads us to prefer a higher value for the initial
jet energy $m_f$. 

\begin{figure}[htbp]
\includegraphics[scale=0.7]{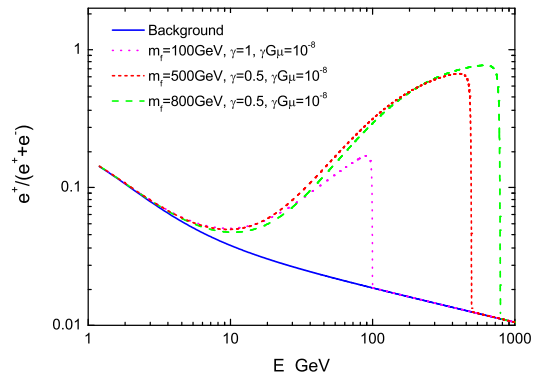}
\caption{Predictions for the flux ratio $\Phi$ at higher energies, for
the same parameter values as in the previous two figures.}
\label{fig:ratio}
\end{figure}

\section{Conclusions and Discussion}

In this Letter we have proposed a new explanation for the observed
excess of positrons over electrons n the cosmic ray flux at energies
between $10$ and $600$GeV. In our model, the source of the
cosmic ray positrons is radiation from cosmic string cusp emission.
Our model does not assume that the particle physics model manifests
supersymmetry with low-scale supersymmetry breaking. In contrast,
it assumes the existence of linear topological defects.

The spectral shape which we predict is insensitive to the details
of the cusp annihilation process and is thus a robust prediction
of our model. The position of the peak of the flux is determined
by the initial jet energy $m_f$. The amplitude of the spectrum, 
is not a robust prediction of our model. It depends sensitively
on both the cosmic string tension and on the efficiency factor $\epsilon$
of the cusp annihilation process. In our plots, we have fixed
$\epsilon = 1$. From the analysis in Section III it follows
immediately that the factor which determines the amplitude
of the flux is $\epsilon (\gamma^{11/6} G \mu)^{-1}$. This is the factor
which can be fixed from the recent positron flux observations,
assuming that our mechanism is the source of the excess.

From our analysis we can learn another important lesson:
for fixed value of $m_f$, a model with non-superconducting
cosmic strings predicts a positron flux with a shape given
by our analysis. Even if the observed flux is {\it not} due
to strings, we get an upper bound on the quantity
$\epsilon (\gamma^{11/6} G \mu)^{-1}$. As our results show,
for $\epsilon \sim 1$ this is a bound which rules out many
models with low energy scale strings.

Let us add some more comments on the sensitivity of our
predictions to the value of the efficiency factor $\epsilon$.
If we were to use to value of $l_c$ given by
\cite{Olum}, which takes into account effects which were not
included in the initial analysis of \cite{RHBcusp}, and is
\be
l_c \, \sim \, w^{1/2} R^{1/2} \, ,
\ee
then the predicted amplitude of the flux decreases by a factor of
$(w/t_0)^{1/6}$. Moreover, back-reaction effects on cusp formation
are still not included completely in \cite{Olum}, and thus the
actual amplitude may even be lower. The assumptions we make
about the efficiency of cusp annihilation will change the value
of $G \mu$ for which the amplitude of the positron flux agrees with
observations.

There are large classes of particle physics models beyond the
Standard Model which predict the existence of cosmic strings.
Cosmic strings are also predicted in many inflationary universe
models based on superstring theory \cite{Tye} (for reviews
see e.g. \cite{BIreview}). A network of
cosmic strings will also remain in the string gas cosmology
model \cite{BV,NBV} (for a recent review see \cite{RHBrev3}).

In order for our model to be consistent with the absence of
an excess in the cosmic ray anti-proton spectrum \cite{Pamela2}
we require the cosmic strings to decay predominantly leptonically.

In light of the recent positron data, our work motivates a closer
look at the mechanism of cusp annihilation. Any improvement
in our understanding of this process would lead to a much improved
predictive power of our analysis. Another issue which merits
re-visiiting is the determination of the initial jet mass $m_f$
resulting from cusp annihilation. Another important issue
is to determine which cosmic string models can lead to
predominantly leptonic jets. On the experimental side, it
is interesting to explore ways to distinguish between the
proposed scenarios to explain the positron excess (see
e.g. \cite{tests}).

\begin{acknowledgments} 
 
We wish to thank  X.-J. Bi, Xue-lei Chen and in particular Yuan Qiang for help and 
useful discussion. RB wishes
to thank the Theory Division of the Institute of High Energy
Physics (IHEP) for their wonderful hospitality and financial support.
RB is also supported by an NSERC Discovery Grant and 
by the Canada Research Chairs Program. The research of Y.C. and X.Z.
is supported  in part by the National Science Foundation of China under Grants 
No. 10533010 and  10675136, by the 973 program No. 2007CB815401, 
and by the Chinese Academy of Sciences under Grant No. KJCX3-SYW-N2
 
\end{acknowledgments}

\end{document}